\documentclass[preprint,12pt]{aastex}
\usepackage{longtable}
\usepackage{graphics}
\usepackage{rotating}
\usepackage{amsmath}
\usepackage{txfonts}
\usepackage{natbib}
\def\gtorder{\mathrel{\raise.3ex\hbox{$>$}\mkern-14mu
             \lower0.6ex\hbox{$\sim$}}}
\def\ltorder{\mathrel{\raise.3ex\hbox{$<$}\mkern-14mu
             \lower0.6ex\hbox{$\sim$}}}

\slugcomment{\today}

\shorttitle{Supernova PTF\,10vgv}

\shortauthors{Corsi et al.}

\begin{document}
 
\title{Evidence for a compact Wolf-Rayet progenitor for the Type Ic supernova PTF\,10vgv}
\author{A.~Corsi\altaffilmark{1}, E.~O.~Ofek\altaffilmark{2}, A.~Gal-Yam\altaffilmark{2},  D.~A.~Frail\altaffilmark{3}, D.~Poznanski\altaffilmark{4}, P.~A.~Mazzali\altaffilmark{5,6}, S.~R.~Kulkarni\altaffilmark{7}, M.~M.~Kasliwal\altaffilmark{7}, I.~Arcavi\altaffilmark{2}, S.~Ben-Ami\altaffilmark{2}, S.~B.~Cenko\altaffilmark{8}, A.~V.~Filippenko\altaffilmark{8}, D.~B.~Fox\altaffilmark{9}, A.~Horesh\altaffilmark{7}, J.~L.~Howell\altaffilmark{9}, I.~K.~W.~Kleiser\altaffilmark{8}, E. Nakar\altaffilmark{10}, I. Rabinak\altaffilmark{2}, R. Sari\altaffilmark{11}, J.~M.~Silverman\altaffilmark{8}, D.~Xu\altaffilmark{2}, J.~S.~Bloom\altaffilmark{8}, N.~M. Law\altaffilmark{12}, P.~E.~Nugent\altaffilmark{8,13}, and R.~M.~Quimby\altaffilmark{7}}

\altaffiltext{1}{LIGO laboratory, California Institute of Technology, MS 100-36, Pasadena, CA 91125, USA; email: corsi@caltech.edu}
\altaffiltext{2}{Department of Particle Physics and Astrophysics, The Weizmann Institute of Science, Rehovot 76100, Israel}
\altaffiltext{3}{National Radio Astronomy Observatory, P.O. Box 0, Socorro, NM 87801, USA}
\altaffiltext{4}{School of Physics and Astronomy, Tel-Aviv University, Tel-Aviv 69978, Israel}
\altaffiltext{5}{INAF-Osservatorio Astronomico, vicolo dell’Osservatorio, 5, I-35122 Padova, Italy}
\altaffiltext{6}{Max-Planck Institut f\"{u}r Astrophysik, Karl-Schwarzschild-Str. 1, D-85748 Garching, Germany}
\altaffiltext{7}{Cahill Center for Astrophysics, California Institute of Technology, Pasadena, CA, 91125, USA}
\altaffiltext{8}{Department of Astronomy, University of California, Berkeley, CA 94720-3411, USA}
\altaffiltext{9}{Department of Astronomy \& Astrophysics, Pennsylvania State University, University Park, Pennsylvania 16802, USA}
\altaffiltext{10}{Raymond and Beverly Sackler School of Physics \& Astronomy, Tel Aviv University, Tel Aviv 69978, Israel}
\altaffiltext{11}{Racah Institute for Physics, The Hebrew University, Jerusalem 91904, Israel}
\altaffiltext{12}{Dunlap Institute for Astronomy and Astrophysics, University of Toronto, 50 St. George Street, Toronto M5S 3H4, Ontario, Canada}
\altaffiltext{13}{Computational Cosmology Center, Lawrence Berkeley National Laboratory, 1 Cyclotron Road, Berkeley, CA 94720, USA}

\begin{abstract}We present the discovery of PTF\,10vgv, a Type Ic supernova detected by the Palomar Transient Factory, using the Palomar 48-inch telescope (P48). $R$-band observations of the PTF\,10vgv field with P48 probe the supernova emission from its very early phases (about two weeks before $R$-band maximum), and set limits on its flux in the week prior to the discovery. Our sensitive upper limits and early detections constrain the post-shock-breakout luminosity of this event. Via comparison to numerical (analytical) models, we derive an upper-limit of $R\lesssim 4.5\,R_{\odot}$ ($R\lesssim 1\,R_{\odot}$) on the radius of the progenitor star, a direct indication in favor of a compact Wolf-Rayet star. Applying a similar analysis to the historical observations of SN\,1994I, yields $R\lesssim 1/4\,R_{\odot}$ for the progenitor radius of this supernova.
\end{abstract}
\keywords{
supernovae: general ---
supernovae: individual (PTF\,10vgv)}

\section{Introduction}
\label{sec:Introduction}%%%%%%%%%%%%%%%%%%%%%%%%
Core-collapse supernovae (SNe) are believed to originate from evolved, massive progenitors (initial mass $\gtrsim 8$--10\,M$_{\odot}$) whose iron core undergoes gravitational collapse. Among them, Type II-Plateau (II-P) SNe show prominent hydrogen in their spectrum and a plateau in the optical light curves. Type IIb SNe have hydrogen in the spectrum initially, and a H-deficient spectrum at later times. Finally, Types Ib and Ic show no evidence for hydrogen at any time. The H-deficient/H-poor core-collapse SNe are thought to be produced by progenitors stripped of their hydrogen (SN~Ib) and possibly helium (SN~Ic) envelopes prior to exploding \citep[for a review, see][]{Filippenko1997}. Due to the stiff dependence of mass loss on luminosity/mass, a sequence of increasing main-sequence mass may be pictured going from progenitors of SNe~II-P, IIb, Ib, and Ic \citep[][]{Heger2003,Crowther2007,Georgy2009}. Rotation, metallicity, and binarity also affect the mass loss \citep[e.g.,][]{Pod1992,Meynet1994,Meynet2000}.

The viability of the standard explosion mechanism for stars of increasing mass is challenging, given that their higher mass cores are more bound, and their SN shocks subject to a very high accretion rate \citep[e.g.,][]{Burrows2007}. Binary-star evolution has been studied as a channel to circumvent this caveat \citep[e.g.,][]{Utrobin1994,Woosley1994,Fryer2007,Yoon2010,Smith2011}. The basic ingredient of this scenario is mass loss through transfer onto a companion. In this case, the mass-loss luminosity scaling does not apply, and much lower mass progenitors can explode as H-poor cores.

Recently, \citet{Dessart2011} published simulations of SN light curves resulting from explosions of SN~IIb/Ib/Ic progenitors. All SNe show a $\sim 10$-day-long post-breakout plateau with a luminosity of $(1-5)\times10^7$\,L$_{\odot}$. Analytical estimates for the early-time ($t\lesssim 1-2$\,d since explosion) post-breakout emission have been provided by \citet{Rabinak2011} and \citet{Nakar2010}. 

In this Letter, we present the discovery of a type Ic SN, PTF\,10vgv, detected by the Palomar Transient Factory\footnote{http://www.astro.caltech.edu/ptf/} \citep[PTF;][]{Law2009,Rau2009}  (\S 2). We report its spectral classification (\S 3) and the radio follow-up observations (\S 4). We constrain the radius of the stellar progenitor of this SN by comparing our tight pre-discovery upper-limits with the predictions of several models \citep{Dessart2011,Rabinak2011,Nakar2010} (\S 5). 
%%%%%%%%%%%%%%%%%%%%%%%%%%%%%%%
\section{Discovery and $R$-band photometry}
\label{sec:Observations}
\begin{figure}
\begin{center}
\includegraphics[height=6.cm]{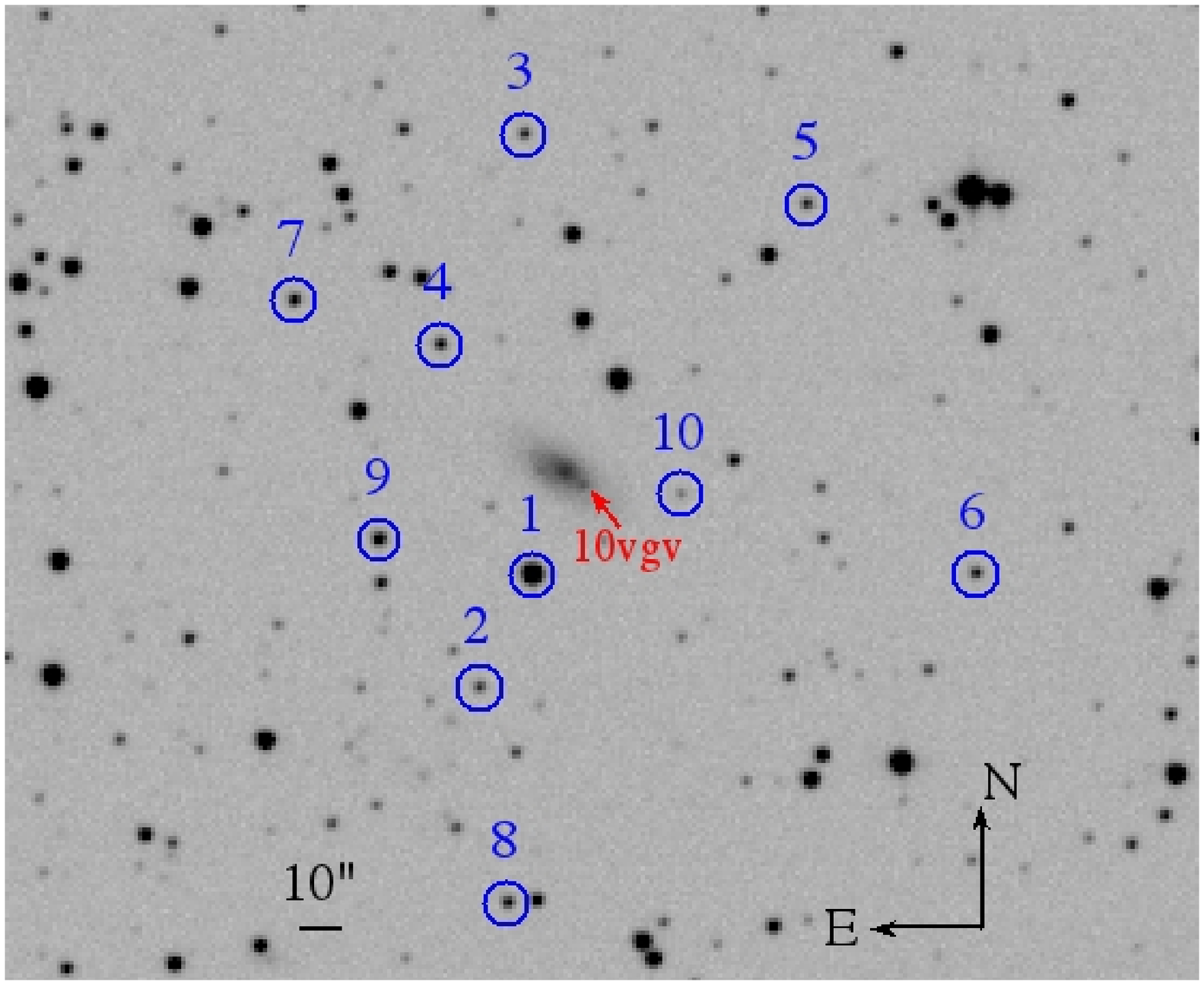}
\caption{Discovery image of PTF\,10vgv (marked with a red arrow) in the $R$ band; the host galaxy is also visible. Circles of $5''$ radius mark the positions of the ten reference stars used for calibration of the P48 photometry (see text). \label{zoom}}
\end{center}
\end{figure}

\begin{figure}
\begin{center}
\includegraphics[width=8.cm]{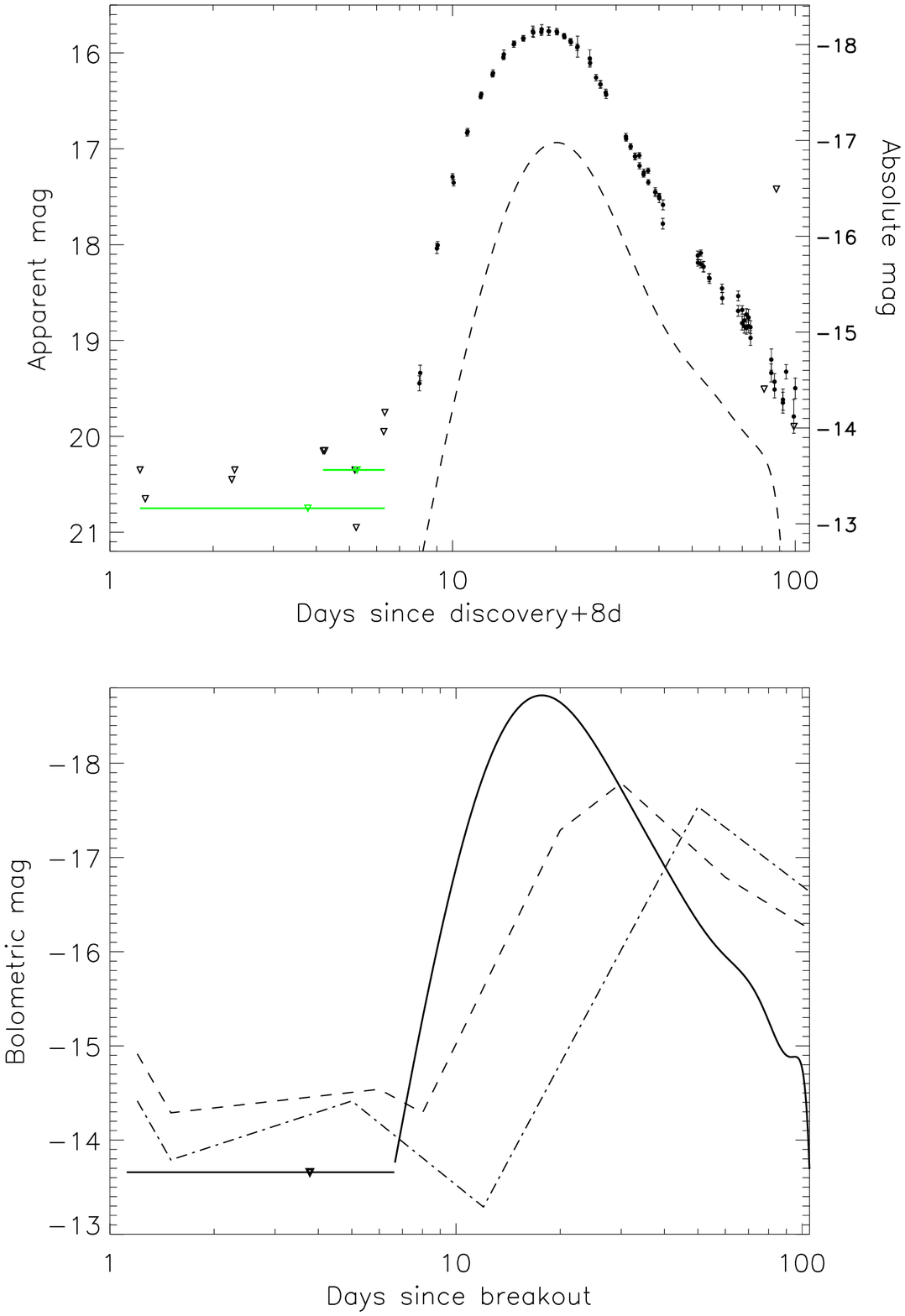}
\caption{\textit{Top:} P48 $R$-band light curve of PTF\,10vgv (black dots) corrected for Galactic extinction. P48 pre-discovery upper limits derived using 60\,s exposure images are plotted as black triangles. Deeper upper limits obtained by coadding the pre-explosion images are plotted as green triangles, with the green horizontal lines indicating the time range spanned by the coadded images. For comparison, we also plot the light curve of SN\,1994I (dashed line), rescaled to the redshift of PTF\,10vgv. \textit{Bottom:} Schematic representations of the bolometric light curves of models Bmi18mf3p79z=1 (dashed line) and Bmi25mf5p09z1 (dash-dotted line) of \citet[][]{Dessart2011} are compared with the PTF\,10vgv bolometric light curve (solid line). The black triangle and solid horizontal line indicate our deepest pre-explosion coadded upper limit (see upper panel), rescaled to account for the bolometric correction (and for Galactic extinction). See \S 5 for discussion.
\label{Fig2}}
\end{center}
\end{figure}

On 2010 September 14.1446 (UTC times are used throughout), we discovered a Type Ic SN, PTF\,10vgv, via the automated Oarical software \citep{Bloom2011}. The SN was visible at a magnitude of $R \approx 19.9$ (Table 1 and Figure \ref{zoom}) in an image (60\,s exposure) taken with the Palomar Oschin Schmidt 48-inch telescope (P48). It was not seen in previous images of the same field taken on 2010 September 12.4830, down to a limiting magnitude of $R > 20.2$. The SN J2000 position is $\alpha = 22^{\rm h}16^{\rm m}01.17^{\rm s}$, $\delta = +40^\circ52'03.3''$ \citep{ATEL2914}, at an angular distance of $\sim 5''$ from the galaxy SDSS\footnote{Sloan Digital Sky Survey \citep{York2000}.} J221601.54+405206.5. 
P48 observations were obtained with the Mould-$R$ filter (Table 1 and Figure \ref{Fig2}). A high-quality image produced by stacking several images of the same field was used as a reference and subtracted from the individual images. Photometry was performed with an aperture of $2''$ radius relative to the $r$-band magnitudes of ten SDSS reference stars in the field (Figure \ref{zoom}), applying color corrections \citep{Corsi2011}. Aperture corrections were applied to account for systematic errors as well as errors introduced by the subtraction process \citep{Corsi2011}.
%%%%%%%%%%%%%%%%%%%%%%%%%%%%%%%%%%%%
\begingroup
\renewcommand{\thefootnote}{\alph{footnote}}
\begin{longtable}{ccc}
\caption{P48 observations of PTF\,10vgv in $R$-band. This Table is published in its entirety in the electronic edition of this journal.\label{Tab1}}\\
\hline
Start time & Exposure & Mag\tablenotemark{a}\\
JD-2455453.6446 (d) & (s)& [mag]\\
\hline
%\startdata
-6.776 & 600 &       $<21.2$\tablenotemark{b}\\
-6.776 & 60 &       $<20.8$\tablenotemark{b}\\
-6.732 & 60 &       $<21.1$\tablenotemark{b}\\
-5.732 & 60 &       $<20.9$\tablenotemark{b}\\
-5.688 & 60 &       $<20.8$\tablenotemark{b}\\
-3.811 & 60 &       $<20.6$\tablenotemark{b}\\
-3.811 & 360 &      $<20.8$\tablenotemark{b}\\
-3.766 & 60 &       $<20.6$\tablenotemark{b}\\
-2.814 & 60 &       $<20.8$\tablenotemark{b}\\
-2.768 & 60 &       $<21.4$\tablenotemark{b}\\
-1.706 & 60 &       $<20.4$\tablenotemark{b}\\
-1.662 & 60 &       $<20.2$\tablenotemark{b}\\
0.000 & 60 &        $19.897\pm0.079$\\
0.044 & 60 &        $19.788\pm0.081$\\
0.996 & 60 &        $18.489\pm0.053$\\
1.043 & 60 &        $18.455\pm0.037$\\
1.994 & 60 &        $17.742\pm0.032$\\
2.068 & 60 &        $17.803\pm0.035$\\
3.019 & 60 &        $17.283\pm0.033$\\
3.064 & 60 &        $17.270\pm0.034$\\
4.075 & 60 &        $16.904\pm0.024$\\
4.122 & 60 &        $16.879\pm0.023$\\
5.073 & 60 &        $16.676\pm0.025$\\
5.117 & 60 &        $16.660\pm0.034$\\
6.065 & 60 &        $16.496\pm0.026$\\
6.108 & 60 &        $16.464\pm0.046$\\
7.078 & 60 &        $16.360\pm0.031$\\
7.122 & 60 &        $16.349\pm0.025$\\
8.058 & 60 &        $16.297\pm0.025$\\
8.106 & 60 &        $16.297\pm0.030$\\
9.148 & 60 &        $16.224\pm0.054$\\
9.193 & 60 &        $16.238\pm0.049$\\
10.149 & 60 &       $16.230\pm0.039$\\
10.193 & 60 &       $16.204\pm0.050$\\
11.060 & 60 &       $16.222\pm0.045$\\
11.103 & 60 &       $16.223\pm0.042$\\
\hline
\footnotetext[1]{Magnitudes are not corrected for Galactic extinction, and are calibrated to the SDSS $r$ (SDSS is estimated to be on the AB system within $\pm 0.01$\,mag in the $r$ and $i$ bands).}
\footnotetext[2]{$3\sigma$ upper limit computed by simulating stars at the position of PTF\,10vgv, to account for the presence of the underlying host galaxy.}
%\footnotetext[3]{$3\sigma$ upper limit.}
\end{longtable}
\endgroup
\renewcommand{\thefootnote}{\arabic{footnote}}
%%%%%%%%%%%%%%%%%%%%%%%%%%%
\section{Spectral classification}
\label{spettrale}
%%%%%%%%%%%%%%%%%%%%%%%%%%%
\begin{figure}
\begin{center}
\includegraphics[height=6.cm]{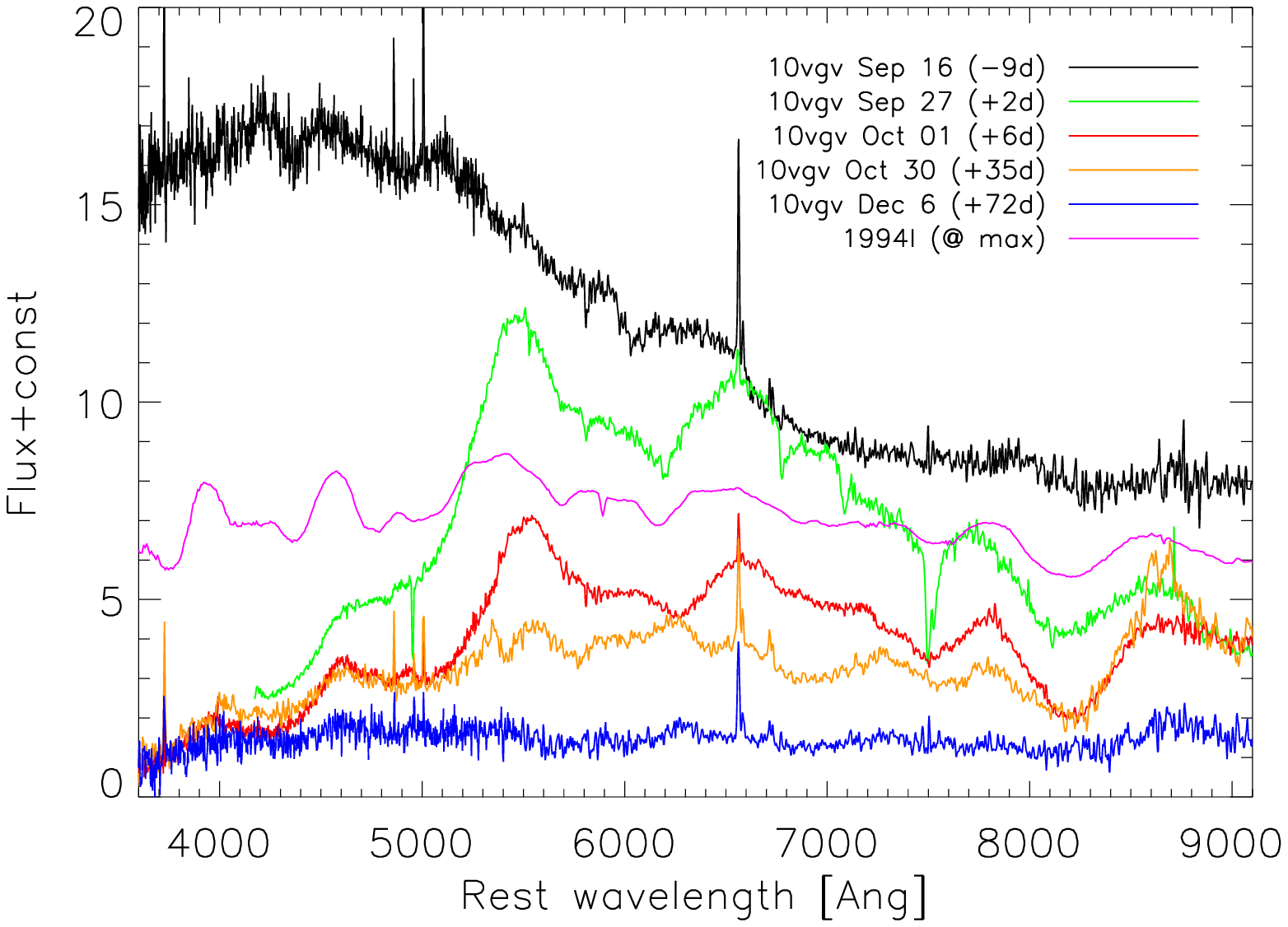}
\caption{Spectra of PTF\,10vgv from Lick/Kast (black and red lines), HET/LRS (green line; telluric absorption lines not removed), and P200/DBSP (orange and blue lines). The approximate epoch since the $R$-band peak is also indicated for each spectrum. For comparison, the spectrum of SN\,1994I around maximum light is also shown (magenta). All data are available in digital form from the Weizmann Institute of Science Experimental Astrophysics Spectroscopy System at http:$//$www.weizmann.ac.il$/$astrophysics$/$wiseass$/$. \label{spectra}}
\end{center}
\end{figure}
After rapidly identifying PTF\,10vgv, we triggered our follow-up programs \citep{Gal-Yam2011}.
On 2010 September 16 and October 1, we observed PTF\,10vgv with the dual-arm Kast spectrograph \citep{ms93} on the 3\,m Shane telescope at Lick Observatory (Figure \ref{spectra}).  We used a $2''$ wide slit, a 600/4310 grism on the blue side, and a 300/7500 grating on the red side, yielding full width at half-maximum intensity (FWHM) resolutions of $\sim 4$\,\AA\ and $\sim 10$\,\AA, respectively.  All observations were aligned along the parallactic angle to reduce differential light losses \citep{f82}. Respective exposure times and air masses were 1800\,s and 1.03 for the first epoch, and 2100\,s and 1.00 for the second epoch. The spectra were reduced using standard techniques \citep[e.g.,][]{fps+03} based on \texttt{IRAF} and  \texttt{IDL} routines. Using the Kast spectra we derive a redshift of $z=0.0142\pm0.0002$ (using H$\beta$, \ion{O}{3}, H$\alpha$, \ion{N}{2}, and \ion{S}{2} lines) for PTF\,10vgv. 

On 2010 September 27, in between the two epochs of the Kast observations, we observed PTF\,10vgv with the Low Resolution Spectrograph (LRS) mounted on the Hobby-Eberly Telescope (HET), using the gr300 grating and GG385 filter.  We applied bias- and flat-field corrections using daytime calibration frames, and removed cosmic rays using the \texttt{IRAF} task ``L.A. Cosmic'' \citep{vandokkum01}. The spectrum was extracted and wavelength-calibrated using the ``apall'' and ``identify'' \texttt{IRAF} tasks, respectively, and had exposure time 450\,s at a mean airmass of 1.26.

On 2010 October 30 and December 07, we observed PTF\,10vgv with the Double Beam Spectrograph \citep[DBSP;][]{Oke1982} on the Palomar 200-inch telescope (P200; Figure \ref{spectra}). We used the 600/4000 and the 158/7500 gratings for the blue and red cameras, respectively, with a D55 dichroic, resulting in a spectral coverage of $\sim$ 3500--9500\,\AA. The spectra were reduced using a custom pipeline combining \texttt{IRAF} and \texttt{IDL} scripts. Respective exposure times and air masses were 600\,s and 1.1 for the first epoch, and 350\,s and 1.04 for the second epoch. 

We measured the velocity of the \ion{Si}{2} absorption at 6355\,\AA, which traces reasonably closely the position of the photosphere \citep[e.g.,][]{Tanaka2008}, using the spectra of PTF\,10vgv taken on September 16, September 27, and October 1. The velocities are $16\times10^{3}$\,km\,s$^{-1}$, $9\times10^{3}$\,km\,s$^{-1}$, and $6\times10^{3}$\,km\,s$^{-1}$, respectively, for the three epochs. These are comparable to those of the ``normal'' SN\,Ic\,1994I at similar epochs \citep[][]{Sauer2006}, $\sim 0.7$ times those measured for SN\,2006aj \citep[associated with X-ray flash 060218;][]{Mazzali2006a}, and smaller than those of the gamma-ray burst (GRB)-associated SN\, 1998bw \citep{Iwamoto1998} and SN\,2003dh (see Figure 5 in Corsi et al. 2011 and references therein). The broad-line SN\,Ic\,2002ap also showed higher velocities \citep[$\gtrsim 16\times10^{3}$\,km\,s$^{-1}$ at $\sim 1$ week after the explosion;][]{GalYam2002,Mazzali2002}.

We thus classify PTF\,10vgv as a normal Type Ic SN. A cursory examination of PTF\,10vgv spectra suggests that the blending of lines in this SN is stronger than in both SN\,2006aj and SN\,1994I, indicating that in PTF\,10vgv there may be significantly more mass at v$\approx2\times10^4$\,km\,s$^{-1}$. In Figure \ref{spectra}, we compare our spectra of PTF\,10vgv with the one of SN\,1994I around maximum light. 
%%%%%%%%%%%%%%%%%%%%%%%%%%%
\section{Radio Follow-up Observations}
Starting on 2010 October 7.16, we observed the position of PTF\,10vgv (along with the necessary calibrators) with the Expanded Very Large Array \citep[EVLA;][]{Perley2009} in its C configuration, at 4.495\,GHz and 7.915\,GHz, for a total time of 30 min \citep{Corsi2010}. We detected no radio emission from the position of PTF\,10vgv, down to 3$\sigma$ limits of $120\,\mu$Jy at 4.495\,GHz and $102\,\mu$Jy at 7.915\,GHz. Based on this, we estimate the 5\,GHz spectral luminosity of PTF\,10vgv to be $\lesssim 5\times10^{26}$\,erg\,s$^{-1}$\,Hz$^{-1}$, or $\sim 100$ times below the radio luminosity of the GRB-associated SN\,1998bw \citep{Kulkarni1998} on a similar timescale. This supports the idea that PTF\,10vgv is a normal SN~Ic, rather than a GRB-associated SN. We reobserved PTF\,10vgv with the EVLA in its BnA configuration starting on 2011 May 12.52, for a total time of 1\,hr and at a central frequency of 8.46 GHz. No radio sources were detected in the error circle of PTF\,10vgv down to a 3$\sigma$ limit of $30\,\mu$Jy. EVLA data were reduced and imaged using the AIPS software package.
%%%%%%%%%%%%%%%%%%%%%%%%%%
\section{Discussion}
The measured peak magnitude of PTF\,10vgv (see Table 1) corrected for Galactic extinction \citep[$A_R \approx 0.45$\,mag;][]{Schlegel1998} gives $M_R = -18.16\pm0.05$ mag. The peak absolute magnitude of SN\,1994I was $M_R = -17.99\pm0.48$ \citep[][]{Richmond1996}, while SN\,2006aj had $M_R = -18.81\pm0.06$ mag \citep{Mazzali2006a}. Since PTF\,10vgv is intermediate, in terms of $R$-band peak luminosity, between SN\,1994I and SN\,2006aj, we estimate its nickel mass $M_{^{56}\rm Ni, 10vgv}$ by interpolating between these two SNe \citep{Sauer2006,Mazzali2006a}, using the scaling $L_{\rm peak}\propto M_{\rm Ni} \tau_c^{-1}$  for the peak luminosity \citep[where $\tau_c$ is the light-curve peak width;][]{Arnett1982}, and considering that the PTF\,10vgv light curve is a factor of $\sim 1.25$ broader than that of SN\,1994I (while we take the same $\tau_c$ for PTF\,10vgv and SN\,2006aj). This yields $M_{^{56}\rm Ni, 10vgv} \approx 0.12$\,M$_{\odot}$. 

The mass and kinetic energy of the SN ejecta scale as \citep{Arnett1982} $M_{\rm ej} \propto \tau^2_c {\rm v_{ph}}$ and $E_{\rm K} \propto \tau^2_c {\rm v^3_{ph}}~$, where ${\rm v_{ph}}$ is the photospheric velocity. Using these scalings, and considering that the photospheric velocities of PTF\,10vgv are comparable to those of SN\,1994I and $\sim 0.7$ times those of SN\,2006aj (\S \ref{spettrale}), we estimate the ejecta mass and kinetic energy of PTF\,10vgv interpolating between SN\,2006aj \citep{Mazzali2006a} and SN\,1994I \citep{Sauer2006}. We get $M_{\rm ej, 10vgv}=(1.5\pm0.3)$\,M$_{\odot}$ and $E_{\rm K, 10vgv}=(0.9\pm0.3)\times10^{51}$\,erg. This estimate may be refined through spectral modeling. Our spectral analysis suggests that a different mass-velocity distribution may be realized in PTF\,10vgv (\S \ref{spettrale}), which may lead to a larger $E_{\rm K}$ than estimated on the basis of the light-curve properties, since the latter are mostly determined by the opacity in the inner ejecta.

Our pre-discovery upper limits can be used to constrain the radius of the stellar progenitor of PTF\,10vgv via comparison with model predictions \citep{Dessart2011,Rabinak2011,Nakar2010}. We apply a bolometric correction to our $R$-band data, computed assuming that the SN emits as a black body at temperature $T_{\rm phot}$ (and neglecting redshift corrections):
\begin{eqnarray} 
\nonumber M_{\rm bol}-M_R = -2.5\,{\rm log}_{10}\left(\frac{4\pi\,(10\,{\rm pc})^2 F_0\int^{\nu_2}_{\nu_1}S(\nu)d\nu}{{\rm L}_{\odot}}\right)+M_{\rm bol, \odot}\\+2.5\,{\rm log}_{10}\left(\frac{\int^{\nu_2/kT}_{\nu_1/kT}S(x)x^3(e^{x}-1)^{-1}dx}{\pi^4/15}\right),~~~\label{bol}
\end{eqnarray}
where %L$_{\odot}=3.89\times10^{33}$\,erg\,s$^{-1}$; 
$M_{\rm bol, \odot}=4.72$; $S(\nu)$ is the P48 Mould-$R$ filter transmission; $\nu_1-\nu_2=(4.1-5.3)\times10^{14}$\,Hz; and $F_{0}$ is the photometric zero-point flux ($F_0=3.631\times10^{-20}$\,erg\,cm$^{-2}$s$^{-1}$Hz$^{-1}$ for AB magnitudes). We conservatively maximize the bolometric correction setting $T_{\rm phot}\approx 10^4$\,K, the largest early-time ($t \lesssim 10$\,d since breakout) temperature predicted by the $^{56}$Ni-rich models of Dessart et al. (2011; see their Figure 2, bottom-left panel). In this way we get $M_{\rm bol}-M_R =-0.496$\,mag. 

The optical luminosity of core-collapse SNe after breakout depends on the ejecta composition (via the opacity parameter), the stellar radius, and the $E_{\rm K}/M_{\rm ej}$ ratio. A larger $E_{\rm K}/M_{\rm ej}$ ratio and a lower He fraction both increase the predicted luminosity, for a given stellar radius \citep[][Equations (25) and (29)]{Rabinak2011}. 

In recent numerical simulations of core-collapse explosions of single and binary progenitors of SNe~IIb/Ib/Ic, \citet{Dessart2011} predicted the existence a $\sim 10$-day-long ($\sim 10$ times shorter than in SNe~II-P) post-breakout\footnote{The breakout of a shock through the stellar surface is predicted to be the first electromagnetic signal marking the birth of a SN \citep[e.g.,][]{Arnett1977,Falk1978,Klein1978,Chevalier1992,Waxman2007,Nakar2010,Rabinak2011}.} plateau, with a luminosity of $(1-5)\times10^7$\,L$_{\odot}$ ($\sim 10$ times smaller than in SNe~II-P). This plateau has the same origin as that observed in SNe~II-P\footnote{The plateau is associated with a cooling and recombination wave (CRW) propagating downward through the SN envelope, separating almost recombined outer layers from strongly ionized inner ones \citep[e.g.,][]{Nadyozhin2003}. During the plateau phase, the photosphere sits on the upper edge of the CRW front, whose downward speed is approximately equal to the outward expansion velocity, thus $R_{\rm phot}\approx$\,const. Since also $T_{\rm phot}\approx$\,const\,$\approx T_{\rm recomb}$ (where $T_{\rm recomb}$ is the recombination temperature), a plateau in the luminosity is expected.}, but in the case of SNe~IIb/Ib/Ic it is predicted to have a smaller duration and luminosity because of a more compact progenitor. 

For PTF\,10vgv we can exclude the presence of a post-breakout plateau with luminosity greater than the one of the compact progenitor model Bmi25mf5p09z1 (Figure \ref{Fig2}, lower panel). We thus derive $R\lesssim 4.4\,{\rm R}_{\odot}$ for the radius of PTF\,10vgv progenitor. However, the stellar models analyzed by \citet{Dessart2011} have $E_{\rm K}/M_{\rm ej}$ lower than we derive here, and a high surface He fraction. So the bound on the progenitor radius derived from the comparison with these models is likely over-estimated.  

Similar limits ($R\lesssim5\,{\rm R}_{\odot}$) can be derived using the predictions by Nakar \& Sari (2010; black line in their Figure 3). But this model is accurate only up to $\lesssim 11$\,hr after the explosion, since recombination is not treated. 

Using $M_{\rm ej, 10vgv}=(1.5\pm0.3)$\,M$_{\odot}$ and $E_{\rm K, 10vgv}=(0.9\pm0.3)\times10^{51}$\,erg, as derived above, the tightest constraint, $ R \lesssim 0.7\,{\rm R}_{\odot}$, is obtained from the C/O model of \citet{Rabinak2011}, that accounts for the dependence of the opacity on the envelope composition. The same model, for an envelope composed of mostly He, gives us $ R \lesssim 1.3\,{\rm R}_{\odot}$. Thus, $ R \lesssim 1\,{\rm R}_{\odot}$ is a reasonable estimate \citep[considering that progenitors of type Ic SNe may contain a small fraction of He in the outer layers;][]{Georgy2009}.

Applying this same analytical model to the first clear detection of SN\,1994I (Sauer et al. 2006, Figure 8; Richmond et al. 1996, Figure 7), we get $R\lesssim 1/4\,R_{\odot}$, %This limit is obtained considering that $R\propto L^{1/0.8}$, where $L$ is the bolometric luminosity \citep[][Equation (29)]{Rabinak2011}, and 
considering that $M_{\rm ej,1994I}\approx M_{\rm ej, 10vgv}$, $E_{\rm K, 1994I}\approx E_{\rm K, 10vgv}$, and that the luminosity of SN\,1994I at the time of detection was $\approx 3$ times smaller than the one of PTF\,10vgv.

Our limits for PTF\,10vgv, $R\lesssim (1-5)\,{\rm R}_{\odot}$, are consistent with a small Wolf-Rayet star \citep[e.g.,][]{Crowther2007}, as expected for a highly stripped SN~Ic. Almost all Galactic WN stars with hydrogen (WNL; e.g., Hamann et al. 2006) have $R\gtrsim 5 R_{\odot}$, and all of those reported there have $R\gtrsim 2 R_{\odot}$. Our result thus favors a progenitor having no hydrogen at the surface (WNE, WC or WO, e.g., Sander et al. 2011), in agreement with the fact that Ic SNe progenitors are generally thought to be stripped of their H- (and He-) rich layers (e.g., Gal-Yam et al. 2005; Smartt 2009).

PTF\,10vgv provides the first constraint on the progenitor radius of a SN ever obtained from optical pre-explosion limits extending up to a week before discovery. Optical surveys with rapid cadence and relatively deep exposures (like PTF) should allow us to study many more objects in this manner. 
\acknowledgments
We thank Boaz Katz and Eli Waxman for useful comments. 
PTF is a collaboration of Caltech, LCOGT, the Weizmann Institute, LBNL,
Oxford, Columbia, IPAC, and UC Berkeley. Staff and computational resources
were provided by NERSC, supported by the DOE Office of Science. Lick
Observatory and the Kast spectrograph are operated by the University of California.
HET and its LRS are supported by UT Austin, the Pennsylvania State University,
Stanford, Ludwig-Maximilians-Universit\"{a}t M\"{u}nchen,
Georg-August-Universit\"{a}t G\"{o}ttingen, and the Instituto de Astronomia
de la Universidad Nacional Autonoma de Mexico. The EVLA
is operated by NRAO for the NSF, under cooperative agreement by
Associated Universities, Inc. We thank the staffs of the above observatories
for their assistance. A.G. and S.R.K. acknowledge support
from the BSF; A.G. further acknowledges support from the ISF,
FP7/IRG, Minerva, the Sieff Foundation, and the
German-Israeli Fund (GIF). A.V.F. and his group 
at UC Berkeley acknowledge generous financial assistance from Gary
\& Cynthia Bengier, the Richard \& Rhoda Goldman Fund,
the TABASGO Foundation, and NSF grant AST-0908886.
A.C. acknowledges support from LIGO, which was constructed by Caltech
 and MIT with funding from the NSF under cooperative agreement PHY-0757058, and partial support from NASA/\textit{Swift} grant NNH10ZDA001N.

\end{document}